\documentclass[12pt]{article}
\usepackage{amsmath,amssymb,bbm}
\usepackage{epsfig} 
\usepackage{cite}

\def\m@th{\mathsurround=0pt }
\def\leftrightarrowfill{$\m@th \mathord\leftarrow \mkern-6mu
	\cleaders\hbox{$\mkern-2mu \mathord- \mkern-2mu$}\hfill
	\mkern-6mu \mathord\rightarrow$}

\def\overleftrightarrow#1{\vbox{\ialign{##\crcr
	\leftrightarrowfill\crcr\noalign{\kern-1pt\nointerlineskip}
	$\hfil\displaystyle{#1}\hfil$\crcr}}}

\newcommand{\be}{\begin{equation}}
\newcommand{\ee}{\end{equation}}

\newcommand{\Tr}{\mathop{\rm Tr}}

\def\shat{\ifmmode \hat{s}\else $\hat{s}$\fi}

\def\UV{\rm UV}
\def\IR{\rm IR}
\def\Le{{\bf L}}
\def\R{{\bf R}}
\def\A{{\bf A}}
\def\J{{\bf J}}

\def\nn{\nonumber}

\newcommand{\newc}{\newcommand}

\newc{\gsim}{\lower.7ex\hbox{$\;\stackrel{\textstyle>}{\sim}\;$}}
\newc{\lsim}{\lower.7ex\hbox{$\;\stackrel{\textstyle<}{\sim}\;$}}
\newc{\ie}{{\it i.e.}}
\newc{\etal}{{\it et al.}}
\newc{\mev}{\hbox{\rm\,MeV}}
\newc{\gev}{\hbox{\rm\,GeV}}
\newc{\tev}{\hbox{\rm\,TeV}}
\newc{\xpb}{\hbox{\rm\, pb}}
\newc{\xfb}{\hbox{\rm\, fb}}

\newc{\G}{{\cal G}}
\newc{\h}{{\cal H}}
\newc{\D}{{\cal D}}
\newc{\E}{{\cal E}}

\newc{\mtop}{m_t}
\newc{\mbot}{m_b}
\newc{\mz}{M_Z}
\newc{\mw}{M_W}
\newc{\alphasmz}{\alpha_s(M_Z)}
\newc{\swsq}{\sin^2\theta_W}
\newc{\cwsq}{\cos^2\theta_W}
\newc{\tw}{\tan\theta_W}
\newc{\cw}{\cos\theta_W}
\newc{\sw}{\sin\theta_W}
\newc{\BR}{\hbox{\rm BR}}
\newc{\zbb}{Z\to b\bar}
\newc{\Gb}{\Gamma (Z\to b\bar b)}
\newc{\Gh}{\Gamma (Z\to \hbox{\rm hadrons})}
\newc{\sgn}{\mbox{sgn}}

\def\I{\rm \mathbbm{1}}

\newcounter{mysubequation}[equation]

\def\beq{\begin{equation}}
\def\eeq{\end{equation}}
\def\bea{\begin{eqnarray}}
\def\eea{\end{eqnarray}}
\def\slashchar#1{\setbox0=\hbox{$#1$}           
   \dimen0=\wd0                                 
   \setbox1=\hbox{/} \dimen1=\wd1               
   \ifdim\dimen0>\dimen1                        
      \rlap{\hbox to \dimen0{\hfil/\hfil}}      
      #1                                        
   \else                                        
      \rlap{\hbox to \dimen1{\hfil$#1$\hfil}}   
      /                                         
   \fi}                                         %
\catcode`@=11
\long\def\@caption#1[#2]#3{\par\addcontentsline{\csname
  ext@#1\endcsname}{#1}{\protect\numberline{\csname
  the#1\endcsname}{\ignorespaces #2}}\begingroup
    \small
    \@parboxrestore
    \@makecaption{\csname fnum@#1\endcsname}{\ignorespaces #3}\par
  \endgroup}
\catcode`@=12

\begin{document}

\baselineskip=18pt

\setcounter{footnote}{0}
\setcounter{figure}{0}
\setcounter{table}{0}

\begin{titlepage}
\begin{flushright}
UAB-FT-650
\end{flushright}
\vspace{.3in}

\begin{center}
{\Large \bf 
Baryon Physics in Holographic QCD
}

\vspace{0.5cm}

{\bf Alex  Pomarol$^{a}$ and Andrea Wulzer$^{b}$}

\vspace{.5cm}

\centerline{$^{a}${\it  IFAE, Universitat Aut\`onoma de Barcelona, 08193 Bellaterra, Barcelona, Spain}}
\centerline{$^{b}${\it Institut de Th\'eorie des Ph\'enom\`enes Physiques, EPFL,  CH--1015 Lausanne, Switzerland}}

\end{center}
\vspace{.8cm}

\begin{abstract}
\medskip
\noindent
In a simple  holographic model for  QCD in which  the Chern-Simons term is incorporated
to take into account  the QCD chiral anomaly,  we show that  baryons arise as 
stable solitons which are the 5D analogs of 4D skyrmions.
Contrary to 4D skyrmions and  previously considered holographic scenarios, these solitons have sizes larger than the inverse  cut-off of the model, and therefore 
they are predictable within our  effective field theory approach.
We perform a numerical  determination 
of  several static properties of the nucleons and find a satisfactory agreement with data.
We also calculate  the amplitudes of  ``anomalous'' processes  induced by the Chern-Simons term  in the meson sector,  such  as $\omega\rightarrow\pi\gamma$ and $\omega\rightarrow 3\pi$. A combined fit to baryonic and mesonic observables leads to an agreement  
with experiments  within $16\%$.
\end{abstract}

\bigskip
\bigskip

\end{titlepage}


\section{Introduction}
\label{intro}
In the large-$N_c$ limit,  strongly interacting theories such as QCD
 have a  dual description in terms of a weakly-interacting  
 theory  of  mesons \cite{'tHooft:1973jz}.
 In this dual description, baryons are expected to appear
 as  solitons  made  of mesons fields,     
 usually referred as skyrmions \cite{Skyrme:1961vq,Adkins:1983ya}.

Skyrmions have been widely studied in the literature, with some phenomenological successes. 
Nevertheless, since the full theory of QCD mesons is not known, 
these studies  have been carried out in truncated low-energy models 
either  incorporating  only pions \cite{Skyrme:1961vq,Adkins:1983ya} or
 few  resonances  \cite{Meissner:1987ge}.
It is unclear whether  these approaches 
capture the physics  needed to fully describe the baryons, since
the stabilization of the baryon size is very sensitive to resonances around the GeV.
In the original Skyrme model with only pions, for instance, the inverse skyrmion size $\rho^{-1}$
equals the chiral perturbation theory cut-off $\Lambda_{\chi PT}\sim4\pi F_\pi$, rendering baryon physics completely 
incalculable. 
Other examples are models with the $\rho$-meson which were shown to  have a stable  
skyrmion solution \cite{Igarashi:1985et}. The inverse size, also in this case, is of order $m_\rho\sim\Lambda_{\chi PT}$, which is clearly not far from the mass of the next resonances.
Including the latter could affect strongly the physics of the skyrmion, or even destabilize it. 
\footnote{In Refs.~\cite{Hata:2007mb,Pomarol:2007kr} it was shown how the inclusion of 
the full  tower of isovector resonances in a $SU(2)_L\times SU(2)_R$ 5D model can destabilize the skyrmion.} 

In this article we will consider a very simple five-dimensional model for QCD, which has already been shown 
to give a quite accurate description of meson physics \cite{Erlich:2005qh,Rold:2005,Hirn:2005nr}. 
This 5D model has a  cut-off scale    $\Lambda_5$ 
which is   above the lowest-resonance mass $m_\rho$. 
The gap among these two scales, which  ensures calculability in the meson sector, 
is related  to  the number of colors $N_c$ of QCD. In the large $N_c$-limit, 
one has $\Lambda_5/m_\rho\rightarrow\infty$ and  the 5D model describes a 
 theory of infinite mesonic resonances.
We will be interested in studying   the  solitons
of this 5D theory that will  correspond to    the  baryons of QCD.
We will find that   these 5D skyrmion-like solitons   have an  inverse size $\rho^{-1}\sim m_\rho$ smaller than the cut-off scale
$\Lambda_5$. 
Therefore, 
contrary to the 4D case, they can be consistently studied with our 5D effective theory.
The expansion parameter which ensures calculability will  be provided by $1/(\rho\Lambda_5)\ll 1$.

Once calculability is established, it is meaningful to compare our predictions with experiments. 
We will study numerically the 5D baryons and calculate several static properties of the nucleons 
such as the axial coupling, magnetic moments and radii.
An important ingredient of our model will be the Chern-Simons (CS) term
that not only  will incorporate the QCD anomaly, but will also 
play a crucial role to stabilize the size of the baryons. Indeed,
without the CS term,  we would be back to the scenario discussed in 
Ref.~\cite{Pomarol:2007kr} in which the skyrmion, though stable and calculable, does not provide a good description of baryons.
This CS term will at the same time be responsible for the anomalous processes of the mesons
such as $\pi\rightarrow\gamma\gamma$, $\omega\rightarrow\pi\gamma$, etc., which we will compute and compare with the experimental data.

We claim that this is the first consistent holographic approach to baryons. Previous studies, although useful to understand the AdS/CFT correspondence for baryon physics,
faced  several problems. The first approaches 
considered only a truncated theory  of resonances   \cite{Nawa:2006gv},  and therefore had 
the same problems as 4D skyrmions. Later studies  
\cite{Hata:2007mb,Hong:2006ta,Hata:2008xc} were performed within the 
Sakai-Sugimoto model \cite{Sakai:2004cn}. It was shown, however,     that baryons are not calculable 
in this framework as their inverse size is of the order of the string scale which 
corresponds to the cut-off of the theory \cite{Hata:2007mb}. 
We will show that in our case we have an expansion parameter  that, although not very small in real QCD ({\it i.e.} for $N_c=3$), allows for a perturbative approach. Once a pertubative series has been formally built up, the accuracy of the predictions one gets at a given order depends on how fast the series converges, and not on the smallness of the naive expansion parameter. As we will see,  the  agreement with data of our leading-order predictions is quite good,   suggesting  that the series could  converge fast in our case. Another important difference of our analysis is that we will find the non-linear soliton solution numerically, without the need of unreliable approximations. 

\section{A five-dimensional  model for QCD}
\label{model}

The 5D model that we will use to describe  QCD with two massless flavors
 is the following.
We will consider an $U(2)_L\times U(2)_R$ gauge theory with  metric  
$ds^2=a(z)^2\left(\eta_{\mu\nu}dx^\mu dx^\nu-dz^2\right)$, where $x^\mu$ represent the usual $4$ coordinates 
 and $z$, which runs in the interval 
$[z_{\UV},z_{\IR}]$,  denotes the extra dimension. 
We will work in AdS$_5$ where the warp factor $a(z)$ is
\begin{equation}
a(z)=\frac{z_{\IR}}{z}\, ,
\end{equation}
and $z_{\UV}\rightarrow 0$ to  be taken at the end of the calculations.
In this limit $z_{\IR}$ coincides with the AdS curvature  and 
the conformal length
\be
  L=\int^{z_{\IR}}_{z_{\UV}} dz\,.
  \ee
We will denote respectively by $\Le_M$ and $\R_M$, where $M=\{\mu,5\}$, 
 the $U(2)_L$ and $U(2)_R$ gauge connections. 
These are parametrized by $\Le_M=L_M^a\sigma_a/2+\widehat{L}_M\I/2$ and $\R_M=R_M^a\sigma_a/2+\widehat{R}_M\I/2$, where $\sigma_a$ are the Pauli matrices. 
The chiral symmetry breaking is imposed on the boundary at $z=z_{\IR}$ (IR-boundary) by the following boundary conditions:
\beq
\left(\Le_\mu-\R_\mu\right)\left|_{z=z_{\IR}}\right.=0\ ,\;\;\;\;\;\;\left(\Le_{\mu 5}+\R_{\mu 5}\right)\left|_{z=z_{\IR}}\right.=0\, ,
\label{irboundary condition}
\eeq
where the 5D field strength is defined as $\Le_{MN}=\partial_M \Le_N-\partial_N \Le_M-i[\Le_M,\,\Le_N]$, and analogously for $\R_{MN}$.
At the other boundary, the UV-boundary, we impose generalized Dirichlet conditions to all the fields:
\beq
 \Le_\mu\left|_{z=z_{\UV}}\right.=\,l_\mu\ , \;\;\;\;\;\; \R_\mu\left|_{z=z_{\UV}}\right.=\,r_\mu \, .
\label{uvboundary condition}
\eeq
We will eventually be interested in taking the 4D ``sources" $l_\mu$ and $r_\mu$ to vanish. 

The AdS/CFT correspondence tells us how to
 interpret  the above  5D model in terms of a 4D
  QCD-like theory, whose fields we will generically denote by $\Psi(x)$ and its action by $S_4$. This is a  strongly coupled 4D theory that possesses an $U(2)_L\times U(2)_R$ global symmetry with  associated Noether currents  $j_{L,R}^\mu$. 
 If the 4D theory were precisely massless QCD with two flavors, the currents would be given by the usual quark bilinear, 
 $\left(j_{L,R}^\mu\right)_{ij}={\overline{Q}}_{L,R}^j\gamma^\mu Q_{L,R}^i$.  Defining $Z[l_\mu,r_\mu]$ as the generating functional of current correlators, the AdS/CFT correspondence   states that \cite{Gubser:1998bc,Witten:1998qj}
\be
Z\left[l_\mu,r_\mu\right]\,\equiv\, 
\int \mathcal{D}\Psi\exp{\left[iS_4\left[\Psi\right]\,+\,i\int d^4 x {\textrm{Tr}}\left(j_{L}^\mu l_\mu+j_{R}^\mu r_\mu\right)\right]}\,=\,
\int \mathcal{D}\Le_M\mathcal{D}\R_M\exp{\left[iS_5\left[\Le,\R\right]\right]}\, ,
\label{corr}
\ee
where the 5D partition function depends on the sources $l_\mu$, $r_\mu$ through 
the UV-boundary conditions in Eq.~(\ref{uvboundary condition}).

The correspondence Eq.~(\ref{corr}) leads to the following implication.
Under local chiral transformations,  $Z$
receives a contribution  from  the $U(2)^3$ anomaly,  which is known in QCD. 
\footnote{We are not considering the $U(1)$-$SU(N_c)^2$ QCD anomaly,
responsible for the $\eta^{\prime}$ mass,
  since this is subleading in the large-$N_c$ expansion.}
This implies \cite{Witten:1998qj,Hill:2006wu,Panico:2007qd} that the 5D action must contain a CS term
\be
S_{CS}\,=\,-i\frac{N_c}{24\pi^2}\int \left[\omega_5(\Le) - \omega_5(\R)\right]\, ,
\ee
whose variation under 5D local transformations which does not reduce to the identity at the 
UV exactly reproduces the anomaly. The CS coefficient, which is anyhow quantized even from a purely 5D point of view, will be fixed to  $N_c=3$ when matching QCD. The CS $5$-form, defining \mbox{$\A=-i\A_Mdx^M$}, is
\be
\omega_5(\A)\,=\,{\rm Tr}\left[\A (d \A)^2+ \frac{3}{2} \A^3 (d \A) + \frac{3}{5} \A^5\right]\, .
\ee
When $\A$  is the connection of an $U(2)$ group, as in our case,
one can use  the fact that $SU(2)$ is an anomaly-free group to write
$\omega_5$  as
\be
\omega_5(\A)\,=\,\frac32{\widehat{A}}{\rm Tr}\left[F^2\right]+\frac14\widehat{A}\left(d\widehat{A}\right)^2\,+\,
d\,{\rm Tr}\left[\widehat{A} A F -\frac14 \widehat{A} A^3\right] \, ,
\ee
where $\A=A+\widehat{A}\I/2$ and $A$ is the $SU(2)$ connection. 
The total derivative part of the above equation  can be dropped, since it only adds to $S_{CS}$ an UV-boundary term for the sources.

The full 5D action will be given by $S_5=S_g+S_{CS}$, where $S_g$ is made of locally gauge invariant terms. $S_g$ is also invariant under transformations which do not reduce to the identity at the UV-boundary,  and for this reason it does not contribute to the anomalous variation of the partition function. Taking the operators of the lowest dimensionality, 
we have
\beq
S_g=-\int d^4{x}\int^{z_{\IR}}_{z_{\UV}} dz\,  a(z)\, \frac{M_5}{2} \left\{
\Tr\left[{L_{MN}L^{MN}}\right]\,+\,\frac{\alpha^2}2{\widehat{L}}_{MN}{\widehat{L}}^{MN}\,+\,\{L\,\leftrightarrow\,R\}\right\}\, .
\label{Sg}
\eeq
We have imposed  on the 5D theory   invariance under the combined $\{{\bf x}\rightarrow-{\bf x},L\,\leftrightarrow\,R\}$, where ${\bf x}$ denotes ordinary $3$-space coordinates. This symmetry, under which $S_{CS}$ is also invariant, corresponds to the usual parity on the 4D side. 
Notice that we 
have normalized differently the kinetic term
of the $SU(2)$ and  $U(1)$ gauge bosons, since 
we do not have any symmetry  reason to put them equal.
In the large-$N_c$ limit of QCD, however, the Zweig's rule leads to equal couplings (and masses) for the $\rho$ and $\omega$ mesons, implying $\alpha=1$ in our 5D model.
Since this well-known feature of large-$N_c$ QCD
 does not arise automatically in our 5D framework (as, for instance, the equality of the $\rho$ and $\omega$ masses does), we will keep $\alpha$ as a free parameter.
The CS term, written in component notation, will be given by
\be
S_{CS}\,=\,\frac{N_c}{16\pi^2}\int d^5x\left\{
\frac14\epsilon^{MNOPQ}{\widehat{L}_M}\Tr\left[L_{NO}L_{PQ}\right]\,+\,
\frac1{24}\epsilon^{MNOPQ}{\widehat{L}_M}{\widehat{L}_{NO}}{\widehat{L}_{PQ}}\,-\,\{L\,\leftrightarrow\,R\}
\right\}\, .
\label{Scs}
\ee
The 5D theory defined above has only 3 independent parameters: 
$M_5$, $L$ and $\alpha$. 

From  Eq.~(\ref{corr})  we can extract the current operators
through which the theory couples to the external EW bosons.
These currents are obtained by varying  Eq.~(\ref{corr}) with respect to  $l_\mu$ (exactly the same would be true for $r_\mu$) and then taking $l_\mu=r_\mu=0$. 
The variation of the l.h.s. of Eq.~(\ref{corr}) simply gives the current correlator of the 4D theory, while in the r.h.s. this corresponds to a   variation of  the UV-boundary conditions. 
The effect of this latter can be calculated in the following way.
We   perform a  field redefinition $\Le_\mu\rightarrow \Le_\mu+\delta \Le_\mu$ where $\delta \Le_\mu(x,z)$  is chosen to  respect the IR-boundary conditions and  
fulfill  $ \delta \Le_\mu(x,z_{\UV})=\delta l_\mu$.  
This redefinition removes the original variation of the UV-boundary conditions, but leads 
a new term  in the  5D action, $\delta S_5$. One then has
\be
i\int d^4x{\textrm{Tr}}\left[\langle j_{L}^\mu(x)\rangle\delta l_\mu(x)\right]\,=\,
i\int \mathcal{D}\Le_M\mathcal{D}\R_M\delta S_5\left[\Le,\R\right]\exp{\left[iS_5\left[\Le,\R\right]\right]}\, ,
\ee
where the 5D path integral is now performed by taking  $l_\mu=r_\mu=0$, {\it{i.e.}} normal Dirichlet 
 conditions. 
 The explicit value  of  $\delta S_5$  is given by
\be
\delta S_5\,=\,\int d^4x {\textrm{Tr}}\left[ \J_{L}^\mu(x)\delta l_\mu(x)\right]\,+\,\int d^5x (\textrm{EOM})\cdot \delta L\, ,
\label{currdef}
\ee
where $\J_{L\, \mu}=J_{L\, \mu}^{a}\sigma^a+\widehat J_{L\, \mu} \I$ and 
\be
J_{L\, \mu}^{a}\,=\,M_5\big(a(z)L_{\mu\,5}^a\big)\left|_{z=z_{\UV}}\right.\ ,
\;\;\;\;\;{\widehat J}_{L\, \mu}\,=\,\alpha^2M_5\big(a(z){\widehat L}_{\mu\,5}\big)\left|_{z=z_{\UV}}\right.\, .
\label{cur0}
\ee
The last term of Eq.~(\ref{currdef}) corresponds to the  5D ``bulk" part of the variation, which leads to the equations of motion (EOM). 
Remembering that the EOM always have zero expectation value \footnote{We have actually shown this here; notice that $\delta \Le_\mu$ was completely arbitrary in the bulk, but the variation of the functional integral can only depend on $\delta l_\mu = \delta \Le_\mu (x,z_{\UV})$.}, 
we find that we can identify $\J_{L}^\mu$ of  Eq.~(\ref{cur0}) with the current operator on the 5D side: $\langle j_L^\mu\rangle_{\rm 4D}=\langle \J_L^\mu\rangle_{\rm 5D}$.
Notice that the CS term has  not contributed to Eq.~(\ref{currdef}) due to the fact that each term in $S_{CS}$ which contains a $\partial_z$ derivative (and therefore could lead to a UV-boundary term) also contains  $\Le_\mu$ or $\R_\mu$  fields;
these fields  on   the UV-boundary  are the sources $l_\mu$ and $r_\mu$ 
that must be put to zero.

\section{Baryons as  5D skyrmions}

\subsection{The static solution}
\label{ss}

The QCD baryons correspond
to the  solitons of the above 5D theory, also referred as 5D skyrmions. 
These static solutions, exactly like the ones considered in Ref.~\cite{Pomarol:2007kr}, have unit 
topological charge 
$$
B=\frac1{32\pi^2}\int d^3x\int^{z_{\IR}}_{z_{\UV}} dz\,  
\epsilon_{\hat\mu\hat\nu\hat\rho\hat\sigma}\Tr\left[{\Le^{\hat\mu\hat\nu}\Le^{\hat\rho\hat\sigma}}
-{\R^{\hat\mu\hat\nu}\R^{\hat\rho\hat\sigma}}\right]\,,
$$
which we identify with the baryon number. 
The derivation of the solitonic configurations   
closely follows the one of Ref.~\cite{Pomarol:2007kr}.
There are, however,   few differences. First, in Ref.~\cite{Pomarol:2007kr} we were using time-reversal symmetry to put consistently to zero the temporal component of the gauge fields. Here, on the contrary, we have to use time-reversal $t\rightarrow-t$ combined with ${\widehat{L}}\rightarrow-{\widehat{L}}$ and ${\widehat{R}}\rightarrow-{\widehat{R}}$, under which  the CS is invariant.  This transformation reduces,
in static configurations,  to a sign change of the temporal component of $L$ and $R$ and of the spatial components of ${\widehat{L}}$ and ${\widehat{R}}$. 
We can therefore consistently put them to zero. 
For the remaining fields one can impose ``cylindrical'' symmetry, {\it i.e.} invariance under combined $SU(2)$ gauge and 3D spacial rotations, and parity invariance under the combined action  $L\leftrightarrow R$ and ${\mathbf x}\rightarrow -{\mathbf x}$. 
This  determines our  ansatz to be 
\be
 L_{i}^a(\mathbf x,z)=-R_{i}^a(-\mathbf x,z)\ ,\ \  L_{5}^a(\mathbf x,z)=R_{5}^a(-\mathbf x,z)\ , \ \  {\widehat{L}}_{0}(\mathbf x,z)={\widehat{R}}_{0}(-\mathbf x,z)\, ,
\label{ans1}
\ee
and
\bea
R^a_j&=&-\frac{1+\phi_2(r,z)}{r^2}\, \epsilon_{jak}{x}_k+\frac{\phi_1(r,z)}{r^3}\left(r^2\delta_{ja}-{x}_j{x}_a\right)+{\frac{A_1(r,z)}{r^2}} {x}_j{ x}_a\,,\nn \\
R^a_5&=&\frac{A_2(r,z)}{r}{ x}^a\,,\nn\\
{\widehat{R}}_0&=&\frac1{\alpha}\frac{s(r,z)}{r}\, .
\label{ans2}
\eea
The above  ansatz  reduces our solitonic configuration
to  $5$ real functions in 2D, $A_{\bar\mu}=\{A_1,A_2\}$, $\phi=\phi_1+i\phi_2$ and $s$,
where  $x^{\bar\mu}=\{r,z\}$.
We notice that we have a residual $U(1)$ invariance corresponding to $g_L^\dagger=g_R={\textrm{exp}}[i\,\alpha(r,z)x^a\sigma_a/(2r)]$ under which $A_{\bar\mu}$ is the gauge field, $\phi$ has charge $+1$ and $s$ is neutral.

The contribution of $S_g$ to the energy  is easily computed. 
The   contribution   from  the $SU(2)$ part is given in
Eq.~(11) of Ref.~\cite{Pomarol:2007kr},  while the $U(1)$ part only adds 
the  kinetic energy of $s$. We then have 
\beq
E_g=8\pi M_5\int_0^\infty dr\int^{z_{\rm IR}}_{z_{\rm UV}}dz\,a(z)\left[
|D_{\bar\mu}\phi|^2+\frac{1}{4}r^2F^{2}_{\bar\mu\bar\nu}
+\frac{1}{2r^2}\left(1-|\phi|^2\right)^2-\frac12\left(\partial_{\bar\mu} s\right)^2\right]\, .
\label{Eg}
\eeq
Notice that $s$ has a negative kinetic term, since it corresponds to the temporal component of a gauge field. The CS term gives also a contribution to the energy.
From the ansatz of Eqs.~(\ref{ans1}) and (\ref{ans2}) we have 
${\widehat{L}}_{0}(r,z)={\widehat{R}}_{0}(r,z)=s/(\alpha r)$,
that allows us to write the CS energy as  
 a coupling of $s$
to the topological charge density (the baryon number density),
given in Eq.~(12) of Ref.~\cite{Pomarol:2007kr}:
\beq
E_{CS}=8\pi M_5\frac{-\gamma L}2\int_0^\infty dr\int^{z_{\IR}}_{z_{\UV}}dz
\frac{s}{r}\epsilon^{\bar\mu\bar\nu}
\bigg[\partial_{\bar\mu}(-i \phi^*D_{\bar\nu}\phi+h.c.)
+F_{\bar\mu\bar\nu}\bigg]\,,
\label{Ecs}
\eeq
where
\be
\gamma=\frac{N_c}{16\pi^2 M_5 L\alpha}\,.
\ee
From Eqs.~(\ref{Eg}) and (\ref{Ecs}) we can  extract the EOM. One finds
\be
\displaystyle
\left\{
\begin{array}{l}
D_{\bar\mu}\Big(a(z)D_{\bar\mu}\phi\Big)+\frac{a(z)}{r^2}\phi\left(1-|\phi|^2\right)+i\,\gamma L\epsilon^{\bar\mu\bar\nu}\partial_{\bar\mu}\left(\frac{s}{r}\right)D_{\bar\nu}\phi=0\\
\partial^{\bar\mu}\Big(r^2a(z)F_{\bar\mu\bar\nu}\Big)-a(z)(i \phi^*D_{\bar\nu}\phi+h.c.)+\gamma L\epsilon^{\bar\mu\bar\nu}\partial_{\bar\mu}\left(\frac{s}{r}\right)\left(|\phi|^2-1\right)=0\\
\partial_{\bar\mu}\Big(a(z)\partial^{\bar\mu} s\Big)-\frac{\gamma L}{2 r}\epsilon^{\bar\mu\bar\nu}
\Big[\partial_{\bar\mu}(-i \phi^*D_{\bar\nu}\phi+h.c.)
+F_{\bar\mu\bar\nu}\Big]=0
\end{array}
\right.\,.
\ee
The skyrmion configurations   will be   the solutions to these EOM
with boundary conditions enforcing a definite  topological charge $B$. 
For the $B=1$ solution, these boundary conditions  
are given in Eqs.~(13) and (14) of Ref.~\cite{Pomarol:2007kr} for the $SU(2)$ fields, while for the neutral scalar $s$ we must impose  $s=0$ at  the three boundaries, $z=z_{\rm UV}$, $r=0$ and $r\rightarrow\infty$,  and impose $\partial_z s=0$ at $z=z_{\rm IR}$. 
The IR and UV-boundary conditions come respectively from Eqs.~(\ref{irboundary condition}) and (\ref{uvboundary condition}); the one at $r=0$ arises from regularity, while the one at $r\rightarrow\infty$ is necessary for the energy to be finite. \footnote{Any constant value for $s$ would give finite energy, but since $s=0$ at the UV, taking $s=0$ is the only possibility.} The solution will be obtained numerically by the COMSOL 3.4 package \cite{comsol}; 
a rescaling of the $\phi_{1,2}$ fields, as explained in Ref.~\cite{Pomarol:2007kr}, must be performed in order to avoid singularities at $r=0$.

At this point it is important to  show that 
the 5D skyrmion configuration  is   stable and its 
properties 
can be
consistently calculated within 
 the 5D effective  theory described above.
This is easily established, along the lines of Refs.~\cite{Hata:2007mb, Pomarol:2007kr},
in the case in which the CS term is a small perturbation to the gauge kinetic term, 
{\it i.e.}, $\gamma\ll 1$.
In this limit, the solution is approximately given by  a small  4D instanton configuration whose
  size  $\rho$ is determined by minimizing  the energy. The AdS curvature
induces  a contribution to the instanton energy that grows as $\rho/L$ \cite{Pomarol:2007kr}, while the 
CS term  generates a Coulomb potential scaling as  $\gamma^2L^2/\rho^2$ \cite{Hata:2007mb}.
Therefore the energy is minimized for  $\rho\sim \gamma^{2/3} L$.
By a Naive Dimensional Analysis (NDA) one gets
\footnote{
In this NDA  we are not considering the  $N_c$ dependence of $\gamma$.
When this is included and  
 our 5D model is matched to large-$N_c$ QCD,  we will see later that
 one gets  $\gamma\sim 1$ and    $\rho\sim L$.
Therefore, including   the $N_c$ factors
will  make the size of the baryon even larger 
 with respect to the cut-off.}
$\Lambda_5\lesssim 24\pi^3 M_5$ and 
 $\gamma\sim  1/(\Lambda_5L)\ll 1$, so we have  that $\rho\gg 1/\Lambda_5$ 
and therefore we expect the soliton 
to be insensitive to higher-dimensional operators.
The situation is quite different in the Sakai-Sugimoto model \cite{Hata:2007mb}.
There the energy  dependence on $\rho$ goes as $E(\rho)\sim \rho^2/L^2+\gamma^2L^2/\rho^2$ 
where the first term comes from the curvature of the 5D space 
and the second from the CS with now
$\gamma\sim 1/(M_{\rm st}L)^2$, being  $M_{\rm st}$  the string-scale.
In this case one gets $\rho\sim \sqrt{\gamma}L\sim 1/M_{\rm st}$;
the size of the soliton is of the order of the inverse  cut-off of the model. This makes baryon physics totally sensitive to string corrections and therefore unpredictable, as already remarked in Ref.~\cite{Hata:2007mb}.

\subsection{Soliton quantization}

Exactly like in the case of the 4D skyrmion \cite{Adkins:1983ya} (see Ref.~\cite{Meissner:1987ge} for a comprehensive review), single-baryon states are described in our model as zero-mode time-dependent fluctuations around the classical static solution. Such zero modes, also called collective coordinates, are associated to the global symmetries of the theory; the situation is similar to that of the kink and the monopole \cite{Weinberg:2006rq}. The global symmetries of our static equations are $U(2)_V$ and $3$-space rotations plus $3$-space translations. The latter are associated with baryons moving with uniform velocity and can be simply ignored if one is only interested in static properties like  magnetic moments, the axial charges and  charge radii. \footnote{These quantities are defined in processes with very low transfer momentum during which the baryon only suffers a negligible acceleration. Our formalism, however, is non-relativistic so that uniform baryon motion cannot be ignored in a generic reference frame. The Breit frame is the correct one to compute form factors \cite{Meissner:1987ge}.} The action of $U(1)_V$ is trivial on all our fields and then we are left with $SU(2)_V$  and $3$-space rotations. The two rotations, however, have the same effect on our cylindrically symmetric solution so that rotating in $3$-space would not lead  to any new configuration which cannot be reached with only $SU(2)$ rotation. Therefore, as in the case of the 4D skyrmion, we only need to consider $3$ collective coordinates which are encoded in a $SU(2)$ matrix $U$.

The zero-modes fluctuations we are interested in are constructed as  follows. Let us perform an $SU(2)_V$ transformation $U$ on the static  solution discussed in the previous section. We obtain
\be
R_{\hat\mu}({\bf x},z;U)\,=\,U\,R_{\hat\mu}({\bf x},z)\,U^\dagger\  ,
\;\;\;\;\;
{\widehat{R}_0}({\bf x},z;U)\,=\,{\widehat{R}_0}({\bf x},z)\,,
\label{eq0m}
\ee
where $\hat\mu=1,2,3,5$; similarly for the $U(2)_L$ gauge fields.
 For a constant $U$, Eq.~(\ref{eq0m})  is also a solution of
the EOM. We now introduce a small 
$t$-dependence  on  $U$, {\it i.e.} a small rotational velocity ${\bf K}^i= -\frac{i}{2}{\rm Tr}[\sigma^iU^\dagger dU/dt]$.
 Eq.~(\ref{eq0m}) is now an infinitesimal  deformation of the static solution. In order for this configuration to describe a zero-mode fluctuation, it should fulfill  the time-dependent EOM 
 to linear order in ${\bf K}$  and with $d {\bf K}/dt=0$. Indeed, zero-modes correspond to directions in  field space  in which uniform and slow motion is permitted.

We therefore need to solve the time-dependent EOM.
 Due to the  invariance under the transformation \mbox{$\{ L\leftrightarrow R, {\bf  x}\leftrightarrow -{\bf x}\}$}, 
 we can restrict the configurations to   $L_i({\bf x},z,t)=-R_i(-{\bf x},z,t)$, $L_{5,0}({\bf  x},z,t)=R_{5,0}(-{\bf x},z,t)$ and analogously for ${\hat L}$, ${\hat  R}$. The EOM for the $R$ fields, after separating temporal from  spatial coordinates, read
\be
\left\{
\begin{array}{l}
D_{\hat\nu}\left(a(z)R^{\hat\nu}_{\;0}\right)+\frac{\gamma  \alpha L}4\epsilon^{\hat\nu\hat\omega\hat\rho\hat\sigma}R_{\hat\nu\hat\omega}{\widehat R}_{\hat\rho\hat\sigma}=0  \\
\alpha\partial_{\hat\nu}\left(a(z){\hat R}^{\hat\nu}_{\;\  0}\right)+\frac{\gamma  L}4\epsilon^{\hat\nu\hat\omega\hat\rho\hat\sigma}\left[{\rm  Tr}\left(R_{\hat\nu\hat\omega}R_{\hat\rho\hat\sigma}\right)  +\frac12{\widehat R}_{\hat\nu\hat\omega}{\widehat  R}_{\hat\rho\hat\sigma}\right]=0\\
D_{\hat\nu}\left(a(z)R^{\hat\nu\hat\mu}\right)-a(z)D_0R_{0}^{\;\  \hat\mu}-\frac{\gamma\alpha   L}2\epsilon^{\hat\mu\hat\nu\hat\rho\hat\sigma}\left[R_{\hat\nu  0}{\widehat R}_{\hat\rho\hat\sigma}+R_{\hat\nu \hat\rho}{\widehat  R}_{\hat\sigma 0}\right]=0\\
\alpha\partial_{\hat\nu}\left(a(z){\widehat  R}^{\hat\nu\hat\mu}\right)-\alpha a(z)\partial_0{\widehat R}_{0}^{\;\  \hat\mu}-\gamma L\epsilon^{\hat\mu\hat\nu\hat\rho\hat\sigma}\left[{\rm  Tr}\left(R_{\hat\nu 0} R_{\hat\rho\hat\sigma}\right)+\frac12{\widehat  R}_{\hat\nu 0}{\widehat R}_{\hat\rho\hat\sigma}\right]=0
\end{array}
\right. \, ,
 \label{eomt}
 \ee
where Euclidean metric is used to rise the spatial indices. We  immediately  see that, once the time-dependent ansatz in  Eq.~(\ref{eq0m}) has been chosen for the fields $R_{\hat\mu}$ and  ${\widehat R}_{0}$, the other components $R_{0}$ and ${\widehat  R}_{\hat\mu}$ cannot be put to zero as in the static case. The  time-dependence of $U$ in Eq.~(\ref{eq0m}) acts as a source for the  latter components, as can be seen by looking at the first and the  fourth EOM. As a result, $R_{0}$ and ${\widehat R}_{\hat\mu}$ must be  turned on. The same situation occurs  in the case of the 4D  skyrmion of Ref.~\cite{Meissner:1987ge} 
in which   the temporal and spatial  components of the $\rho$ and $\omega$ mesons
are turned on in the skyrmion quantization (also in the case of the magnetic monopole
for    the temporal component of the  gauge field).

One can show that the second and  the third EOM of Eq.~(\ref{eomt}) are solved, to linear order in ${\bf K}$,  
by the ansatz in Eq.~(\ref{eq0m}) if the fields $R_{0}$ and ${\widehat R}_{\hat\mu}$ are chosen to be linear in ${\bf K}$. Under this assumption,
   ${\widehat  R}_{\hat\nu 0}$ is equal to $\partial_{\hat\nu }{\widehat{R}_0}$ up to  terms proportional to $d{\bf K}/dt$  or quadratic in ${\bf K}$ that
     can be ignored in the approximation of slow motion discussed above.  
   Furthermore,  $R_{\hat\nu0}$ and ${\widehat R}_{\hat\rho\hat\sigma}$ are  of order ${\bf K}$ and $D_0R_{\hat\mu 0}=0$. 
 Using this, the second and the third EOM of Eq.~(\ref{eomt})
    reduce to the static equations of  sec.~\ref{ss}.

We are left with the first and fourth equations of Eq.~(\ref{eomt}), which are $7$ elliptic
equations for the $7$ fields  $R_0$ and ${\widehat{R}_{\hat\mu}}$; those
are the analog of the Gauss law constraint for dyons in the case of
monopoles \cite{Weinberg:2006rq}. 
To solve such equations we can again make a  2D ansatz
following a  generalization of the cylindrical
symmetry we used in the static case 
 in which the rotational velocity ${\bf K}$ also rotates together with the $3$-coordinates ${\bf x}$. 
The resulting 2D equations can  be  solved numerically, but we leave this
for future work.

The quantization of the collective coordinates, from this point on,
exactly proceeds as for the 4D skyrmion
\cite{Adkins:1983ya,Meissner:1987ge}. First, one
plugs  the zero-mode configuration into the action and obtains a
Lagrangian
$$
{\mathcal L}=-M+\lambda{\rm Tr}\left[\partial_0 U\partial_0
U^\dagger\right]\,,
$$
where $M=E_g+E_{CS}$ is the classical mass of the soliton obtained from the
static solution, while $\lambda$  depends also  on the solutions for
$R_{\hat\mu}$ and ${\widehat{R}_0}$. Now, the collective coordinates $U$ are
treated as quantum mechanical variables,  reducing the problem to the
one of quantizing a spherical rigid rotor with momentum of inertia
$\lambda$. The quantization is therefore performed in a standard way and
the energy eigenstates are interpreted as baryon states. 
One finds, as expected in the large-$N_c$ limit, an infinite tower of baryons 
with increasing spin/isospin. More precisely, baryons are in the $(p/2,p/2)$ representation 
of the spin/isospin $SU(2)\times SU(2)$ group; the first two levels $p=1,2$ are interpreted, respectively, as the nucleons and the $\Delta$ multiplet.
We do not need to
repeat this procedure in detail here, but only give the  following useful relation 
that can be derived from quantization rules \cite{Adkins:1983ya}:
\be
{\textrm{Tr}}\left[U\sigma^bU^\dagger\sigma^a\right]\,=\,-\frac83\,S^b\,I^a\,,
\label{qrule}
\ee
where $S^a$  and $I^a$ are  respectively the spin and isospin operators.

\subsection{Static properties of baryons}

There are   several static baryon observables which
are independent of $R_0$ and ${\widehat{R}_{\hat\mu}}$ and therefore  can
be computed by  knowing only the static solution of sec.~\ref{ss} 
together with the  anstaz in Eq.~(\ref{eq0m}). 
By looking at
Eq.~(\ref{cur0}) we see that the spatial components of the vector and
axial currents, and the temporal component of the scalar one are
independent of $R_0$ and ${\widehat{R}_{\hat\mu}}$ up to order ${\bf K}^2$ or
$d{\bf K}/dt$ terms. By plugging Eq.~(\ref{eq0m}) into these currents one finds
\bea
&&J_{A\, i}\,=\,-\,M_5\,a(z_{\rm UV})\left[\delta_{a\,i}\frac{D_z\phi_1}{r}+\frac{x_ix_a}{r^2}\left(F_{zr}-\frac{D_z\phi_1}{r}\right)\right]_{z_{\rm UV}}\,U\sigma^a\,U^\dagger\,,\nn \\
&&J_{V\, i}\,=\,-\,M_5\,a(z_{\rm UV})\left[\epsilon_{a\,i\,k}x_k\frac{D_z\phi_2}{r^2}\right]_{z_{\rm UV}}\,U\sigma^aU^\dagger\,,\nn \\
&&{\widehat{J}}_{V\,  0}\,=\,-2\,M_5a(z_{\rm UV})\frac{\alpha}{3}\left[\frac{\partial_z
s}{r}\right]_{z_{\rm UV}}\,.
\label{curr}
\eea
The axial current is defined as $J_{A\, \mu}=J_{R\, \mu}-J_{L\, \mu}$, the
vector as $J_{V\, \mu}=J_{R\, \mu}+J_{L\, \mu}$ and the scalar vector current
($1/2$ of the baryon number current) is
${\widehat{J}}_{V\, \mu}=1/3\left({\widehat{J}}_{R\,\mu}+{\widehat{J}}_{L\,\mu}\right)$.
From Eq.~(\ref{curr}), using Eq.~(\ref{qrule}),
we can extract the baryon axial coupling $g_A$, the isovector magnetic moment $\mu_V$,
and the isoscalar electric charge radius $r_{E,S}^2$:
\bea
&&g_A\,=\,-\frac{N_c}{9\pi\alpha\gamma}\,\frac1{L}\int_0^\infty dr\,
r\left[a(z)\left(2D_z\phi_1+rF_{zr}\right)\right]_{z_{\rm UV}}\, ,\nn \\
&&\mu_V\,=\,\frac{N_c\,M_N L}{9\pi\alpha\gamma}\,\frac1{L^2}\int_0^\infty dr\,
r^2\left[a(z)D_z\phi_2\right]_{z_{\rm UV}}\, ,\nn \\
&&r_{E,S}^2\,=\,-\frac{L^2}{\pi\gamma}\,\frac1{L^3}\int_0^\infty dr\,
r^3\left[a(z)\partial_zs\right]_{z_{\rm UV}}\, .
\label{stat}
\eea
For the definition of these observables we follow the conventions used in
Ref.~\cite{Meissner:1987ge}. It should be noticed that Eq.~(\ref{stat}) 
has the right  scaling in $N_c$ \cite{Witten:1979kh} if, in accordance with the AdS/CFT expectation, 
the 5D parameters $\gamma$, $\alpha$ and $L$ scale like ${N_c}^0$. We have numerically calculated all the quantities of Eq.~(\ref{stat}) together  with the 
mass of the baryon which is given by the classical mass $M$.
The 3 parameters of our 5D model, $M_5$, $L$ and $\alpha$, 
 are determined  from (1) the pion decay constant 
$F^2_\pi=8M_5\int dz/a(z)$ \cite{Hirn:2005nr}, (2) the mass of the $\rho$, $m_\rho\simeq 3\pi/(4 L)$ \cite{Erlich:2005qh,Rold:2005}
and (3) the  decay constant ratio   $F_\omega/F_\rho=\alpha$.
The results are given in Table~\ref{preba}
 where we  compare them with the experimental values.
We restricted our comparison to nucleon observables (including the mass, even though all baryons are degenerate at the classical level),
since  we
expect higher spin/isospin states (like the $\Delta$-multiplet) to receive
larger corrections. \footnote{In our model 
 $S^a=\lambda{\bf K}^a/2$.
Since we are
performing a small ${\bf K}$ expansion, 
small spin (and isospin) states are 
predicted more reliably.
}
We recall that these predictions are obtained  in the semiclassical (large-$N_c$) limit, and therefore 
they should be valid  up to corrections $\sim 1/N_c$. Consistently with this picture, Table~\ref{preba} shows agreement with experiments within $30\%$.

\begin{table}[ht]
\small
\begin{center}
\begin{tabular}{ccccc}
\hline
  & {\rm Experiment}
    &{\rm AdS$_5$}
\\ \hline
 $M_N$ (MeV) & $940 $ & $1140$ 
\\
 $\sqrt{\langle r^2_{E,S}\rangle}$ (fm)  & $0.79 $ &  $0.94$
       
\\
 $g_A$ & $1.25 $ & $1.0$
\\
 $\mu_p-\mu_n=2\mu_V$ & $4.7 $ & $3.9$
\\
\hline
\end{tabular}
\caption{Predictions for  the baryon static quantities 
 where $M_5$, $L$ and $\alpha$ have been determined   from the experimental values of 
$F_\pi= 87$ MeV, $m_\rho=775$ MeV and $F_{\omega}/F_\rho=0.88$. }
\label{preba}
\end{center}
\end{table}

In any model with exact spontaneously broken chiral symmetry the axial coupling $g_A$ is related to the pion-nucleon coupling $g_{\pi NN}$ by the Goldberger-Treiman relation
\begin{equation}
g_A=\frac{g_{\pi NN}F_\pi}{M_N}\, .
\end{equation}
This relation holds in our 5D model and can be used  to derive 
$g_{\pi NN}$ from the value of $g_A$ and $M_N$ obtained above. We
get $g_{\pi NN}\simeq 13.1$ that is quite close to the  experimental 
value $g_{\pi NN}|_{\rm exp}\simeq 13.5$.

\section{Meson anomalous couplings  from the CS term}

The CS term of Eq.~(\ref{Scs}) is responsible for the anomalous couplings
of the  mesons to the photon and among themselves. 
\footnote{For a study of these processes in the Sakai-Sugimoto model see Ref.~\cite{Sakai:2005yt}.}
In order to compute amplitudes involving a (real or virtual) photon at leading order in the electric 
charge $e$, one could proceed ``holographically'' and use directly Eq.~(\ref{corr}) and Eq.~(\ref{cur0}) 
 to compute matrix elements of the electromagnetic current $J_{em}^\mu={\hat J}_{V}^\mu/3+J_{V}^{3\,\mu}$.
Looking at the explicit form of this current, one easily realizes that electromagnetic interactions only 
proceed, in this ``holographic basis'', through the single exchange of  resonances.  
Our model implements the Vector Meson Dominance (VMD) hypothesis, in which the
photon only couples to hadrons by the mixing with  the $\rho$ and the $\omega$.

A second way to perform the same calculations, which in some cases is simpler, 
is to use a Kaluza-Klein (KK) expansion of the 5D gauge fields, in which the photon mixing matrix is
 automatically diagonalized. This can be done by giving a dynamics to the photon, which is the source 
associated to the electromagnetic current in Eq.~(\ref{corr}). Making the source dynamical means 
integrating also over it in the path integral and this is the same as changing from Dirichlet to Neumann the UV-boundary condition of the corresponding 5D field ${\widehat V}_{\mu}/3+V_{\mu}^3$, where $V=L+R$. The electric charge is fixed to its experimental value $e$ by 
adding a localized kinetic term for the photon of the form $-\left(1/e^2-{\rm c.t.}\right)F^2/4$ 
where the ``counterterms'' c.t. are needed to cancel the charge renormalization induced by QCD ({\it i.e.} bulk) effects. The latter are finite as long as $z_{\UV}$ is finite. At leading order in the charge $e$, 
however, the localized kinetic term is infinite (as $e\rightarrow 0$), but the KK decomposition in the presence of a 
localized kinetic term is well known. One has a massless zero-mode with flat wave function for  $\hat V_\mu/3+V^{3}_\mu$, that we  
identify with the ``diagonalized'' photon $A^{(\gamma)}_\mu$, which is now a mass eigenstate. 
One also has massive KK's which obey, in the limit $e\rightarrow0$, Dirichlet conditions at the UV. 
One basically 
returns to the original theory with Dirichlet 5D fields, but with an extra photon zero-mode $A^{(\gamma)}_\mu$ which  decouples as $e\rightarrow0$. The other relevant states
are the pions $\pi^a$, which are identified with
the zero modes of the fifth component of the axial gauge bosons, 
and  the  $\rho$ and $\omega$
resonances which are respectively  the first isosinglet and isotriplet vector  KK-states.
We  then have
\begin{eqnarray}
A^a_5(x,z)&=&\frac{1}{\sqrt{M_5L}}f_\pi(z)\, \pi^a(x)+\dots\, ,\nonumber\\
\hat V_\mu(x,z)&=&\frac{\sqrt{2}}{3}A^{(\gamma)}_\mu+\frac{1}{\alpha 
\sqrt{M_5L}}f_V(z)\, \omega_\mu+\dots\, ,\nonumber\\
V^{3}_\mu(x,z)&=&{\sqrt{2}}A^{(\gamma)}_\mu+\frac{1}{\sqrt{M_5L}}f_V(z)\, 
\rho_\mu+\dots\, ,
\label{kkexp}
\end{eqnarray}
where $f_\pi(z)$  and $f_V(z)$
are respectively  the 5D wave-function of the pions and vectors. We have
  $f_\pi(z)=1/(a(z)N_\pi)$ and, for the AdS metric,
  $f_V=z J_1(M_Vz)/(N_VL)$ with $M_V\simeq 3\pi/(4L)\simeq m_\rho=
m_\omega$
  where in both cases  $N^2_i=\int dz\, a(z) f^2_i(z)/L$.
Inserting Eq.~(\ref{kkexp}) into the CS term, we obtain the couplings 
between the pions  and two vectors:
\begin{eqnarray}
{\cal L}_{\pi VV}&=&-\frac{N_c}{48\pi^2}\frac{1}{F_\pi}\pi^0
F^{(\gamma)}_{\mu\nu}\widetilde  F^{(\gamma)\, \mu\nu}
-\frac{N_c}{16\pi^2}\frac{g_{\rho\pi\pi}}{F_\pi}
\left[ \frac{1}{\alpha}\pi^0 F^{(\omega)\, {\mu\nu}}+\frac{1}{3}\pi^a
  F^{(\rho^a)\, \mu\nu}  \right]\widetilde F^{(\gamma)}_{\mu\nu}\nonumber\\
&-&\frac{N_c}{16\pi^2}\frac{g^2_{\rho\pi\pi}}{F_\pi}\frac{x}{\alpha}\pi^a
F^{(\omega)}_{\mu\nu}\widetilde  F^{(\rho^a)\, \mu\nu}+\dots\, ,
\label{anoma}
\end{eqnarray}
where $\widetilde F^{\mu\nu}=\epsilon^{\mu\nu\rho\sigma}F_{ \rho\sigma}/2$,
\begin{equation}
g_{\rho\pi\pi}=\frac{1}{\sqrt{2M_5L^3}}\int dz\, a(z)f_\pi^2f_V \ \ {\rm 
and}\ \ \ x=\frac{1}{\sqrt{2}M_5L^2g_{\rho\pi\pi}^2}\int
  dz\, f_\pi f^2_V\, .
\end{equation}
The value of $x$   turns to be very close to 1; for AdS we find 
$x\simeq 1.18$.
We will understand later why this is the case. We also want to remark that
Eq.~(\ref{anoma}) shows  an interesting
  relation between the $\omega\gamma\pi$ (and
$\rho\gamma\pi$) coupling and $g_{\rho\pi\pi}$, the coupling of the 
$\rho$  to two pions.
This relation is fulfilled  for  any five-dimensional  space.

 From Eq.~(\ref{anoma}) we can calculate several meson partial decay widths.
The first term of Eq.~(\ref{anoma}) leads to the decay of the $\pi^0$ to 
two photons
in accordance with the anomaly prediction.
The  decay widths $\Gamma(\omega\rightarrow \pi\gamma)$ and
$\Gamma(\rho\rightarrow \pi\gamma)$ arise  from the second term of 
Eq.~(\ref{anoma}),
  while
$\Gamma(\omega\rightarrow 3\pi)$  proceeds through
   virtual rhos,  $\omega\rightarrow\rho^{(n)\, *} \pi\rightarrow 3\pi$.
   This latter process  is dominated by the lowest state, the $\rho$,
  whose  $\omega\rho\pi$ coupling is given by the third term of 
Eq.~(\ref{anoma}):
  \begin{eqnarray}
A[\omega_\mu(p)\rightarrow\pi^0(q_0)+\pi^+(q_+)+\pi^-(q_-)]&=&
\frac{N_c}{4\pi^2}\frac{g_{\rho\pi\pi}^3}{F_\pi 
m^2_\rho}\frac{x}{\alpha}\epsilon_{\mu\nu\rho\sigma}q_0^\nu q_+^\rho 
q_-^\sigma\big[
D((q_++q_-)^2)\nonumber\\
&+&D((q_++q_0)^2)+D((q_-+q_0)^2)\big]\, ,
\end{eqnarray}
where $D(p^2)=m^2_\rho/(m^2_\rho-p^2)$.
The  predictions for these partial decay widths are given in Table~\ref{preano}, showing a very good agreement with the experimental data.

\begin{table}[tp]
\small
\begin{center}
\begin{tabular}{ccccc}
\hline
   & {\rm Experiment}
     &{\rm AdS$_5$}
\\ \hline
  $\Gamma(\omega\rightarrow \pi\gamma)$ & $0.75$   & $0.86$
             \\
  $\Gamma(\omega\rightarrow 3\pi)$ & $7.6$   & $6.1$
             \\
  $\Gamma(\rho\rightarrow \pi\gamma)$ & $0.068$   & $0.072$
  \\
  $\Gamma(\omega\rightarrow \pi\mu\mu)$ & $8.2\cdot 10^{-4}$   & 
$7.9\cdot 10^{-4}$
  \\
  $\Gamma(\omega\rightarrow \pi e e )$ & $6.5\cdot 10^{-3}$   & 
$7.8\cdot 10^{-3}$
            \\
\hline
\end{tabular}
\caption{Prediction of the anomalous partial decay widths in MeV
where $M_5$, $L$ and $\alpha$ have been determined   from the experimental values of 
$F_\pi= 87$ MeV, $m_\rho=775$ MeV and $F_{\omega}/F_\rho=0.88$. }
\label{preano}
\end{center}
\end{table}

The CS term also contributes to different pion form factors.
For calculating   form factors, however,    it is more suitable  to work in the 
 holographic basis.
We have seen that in this basis the model exhibits the property of VMD.
For example, the  decay $\omega\rightarrow \pi\gamma$ proceeds as
$\omega\rightarrow \pi\rho^{(n)}\rightarrow \pi\gamma$ and similarly for 
 rho decays. 
As in  VMD models   \cite{Kaymakcalan:1983qq}, this allows us  to derive the
  following sum-rule that relates  the $\omega\gamma\pi$ coupling, 
$g_{\omega\gamma\pi}$,
and the $\omega\rho\pi$ coupling, $g_{\omega\rho\pi}$:
\begin{equation}
g_{\omega\gamma\pi}=\sum_n\frac{g_{\omega\rho^{(n)}\pi}F^{(n)}_\rho}{m_{\rho^{(n)}}}\simeq
\frac{g_{\omega\rho\pi}F_\rho }{m_\rho}\, ,
\label{sr}
\end{equation}
where $F^{(n)}_\rho$ are the rho's decay constants that can be found  in 
Ref.~\cite{Erlich:2005qh,Rold:2005}.
It is  easy to verify   Eq.~(\ref{sr}).
Using Eq.~(\ref{anoma}),  Eq.~(\ref{sr}) gives 
$xg_{\rho\pi\pi}F_\rho\simeq m_\rho$
that, since $x\simeq 1$, implies  $g_{\rho\pi\pi}F_\rho\simeq m_\rho$.
This latter equation was derived in Ref.~\cite{Rold:2005} valid for any 
five-dimensional model.
We can now easily calculate form factors.
The $\pi^0\gamma\gamma^*$  form factor, $F_{\pi\gamma}(q^2)$,
at low Euclidian momentum is dominated by the exchange of the $\omega$. 
We then have,
as in VMD, $F_{\pi\gamma}(q^2)=m^2_\omega/(m^2_\omega+q^2)$,
where we have normalized  the form factor as $F_{\pi\gamma}(0)=1$.
We obtain
\begin{equation}
F^\prime_{\pi\gamma}(0)=\frac{a}{m^2_\pi}\ ,\ \ \
a=\frac{m^2_\pi}{m^2_{\omega}}\simeq 0.03\, ,
\end{equation}
  in perfect agreement with the experimental value \cite{Yao:2006px}: 
$a|_{\rm exp}\simeq 0.032\pm 0.004$.
This form factor was previously studied in   Ref.~\cite{Grigoryan:2008up}.
We  can also calculate the partial decay width
$\Gamma(\omega\rightarrow \pi\mu\mu (ee))$ that proceeds through  a 
virtual photon.
This process is  proportional to the $\omega\pi\gamma^*$ form factor that
in our model is simply given by
$F_{\omega\pi}(p^2)=A(\omega\rightarrow \pi\gamma)D(p^2)$,
where $A(\omega\rightarrow \pi\gamma)$ is the on-shell
$\omega\rightarrow \pi\gamma$ amplitude.
The prediction  for this partial decay width  is given 
in Table~\ref{preano}.

Let us finalize this section with the following comment.
In five-dimensional models arising from string
theory, the effective gauge theory consists  of  the DBI term and the CS term.
From the DBI term  arises not only
the kinetic term of the gauge bosons, but also higher-dimensional operators 
suppressed by the string scale. 
Since  the anomalous couplings discussed in this section can only arise from the CS and not from  the DBI,  these couplings will  not receive corrections from higher-dimensional operators.

\section{Global fit to mesonic and baryonic  observables}

In the previous sections we have computed
several properties of baryons and mesons,
 and we have shown that they agree with the experimental values within $30\%$. Nevertheless, in order to gain a better insight on the statistical significance 
of this approach, we will perform in this section a combined fit to many (baryonic and mesonic) physical observables. Our list of observables is presented in  Table~\ref{pretotal}; we have   taken the physical quantities calculated in this article, together with  other  mesonic observables calculated in Refs.~\cite{Erlich:2005qh,Rold:2005,Hirn:2005nr,Panico:2007qd}.
Our  global fit is carried out by   minimizing the root mean square (RMS) error defined as
\be
{\rm RMSE}(M_5,L,\alpha)=\sqrt{\frac{1}{N_{\rm p}}\sum_i \frac{({\cal O}^i_{\rm pre}-{\cal O}^i_{\rm exp})^2}{{\cal O}^{i\, 2}_{\rm exp}}}\, ,
\label{rmse}
\ee
where ${\cal O}_{\rm pre}^i$ denotes  the predictions of our model for any observable,
 ${\cal O}_{\rm exp}^i$ its experimental value and $N_{\rm p}$ the number of predictions minus the number of parameters. In our case $N_{\rm p}=15$.

The statistical meaning of this procedure is the following. All our predictions are obtained at leading order in the $(16\pi^2 M_5L)^{-1}\sim 1/N_c$ expansion, so that quite large quantum corrections are expected, which translate in quite large theoretical errors. Let us assume that all the observables have the same relative theoretical error $\xi$, {\it i.e.} $\Delta{\cal O}^i_{\rm pre}=\xi{\cal O}^i_{\rm exp}$; by looking at Eq.~(\ref{rmse}) one immediately sees that ${\rm RMSE}(M_5,L,\alpha)=\xi \sqrt{\chi^2/N_{\rm p}}$, where $\chi^2$ is the usual chi-squared variable constructed with errors $\Delta{\cal O}^i_{\rm pre}$. 
The minimum of Eq.~(\ref{rmse})  therefore represents the minimal value of the relative corrections $\xi$ for which our model would successfully pass the $\chi^2$ test, {\it i.e.} for which $\chi^2=N_{\rm p}$. All the observables  included in the fit have  an experimental error $\lesssim 10\%$, which we can neglect because it is  smaller, as we will see, than the final  RMS error of the fit.

\begin{table}[tp]
\small
\begin{center}
\begin{tabular}{ccccc}
\hline
  & {\rm Experiment}
    &{\rm AdS$_5$}& Deviation
\\ \hline
 $m_\rho$ & $775 $ & $850$ & $+10\%$
\\
 $m_{a_1}$ & $1230 $ &  $1390$    &$+13\%$
\\
 $m_\omega$ & $782 $ & $850$ &$+9\%$
\\
 $F_\rho$ & $153$ & $175$ &$+14\%$
\\
 $F_{\rho}/F_\omega$ & $0.88$   &$0.90$&$+2\%$
\\ 
 $F_{\pi}$ & $87$ &    $91$& $+5\%$
\\ 
 $g_{\rho\pi\pi}$ & $6.0$ & $5.4$& $-10\%$
\\
 $L_9$ & $6.9\cdot 10^{-3}$ &     $6.2\cdot 10^{-3}$&$-10\%$
\\
 $L_{10}$ & $-5.5\cdot 10^{-3}$ &    $-6.2\cdot 10^{-3}$&$+12\%$
\\
             $M_N$ & $940 $ &  $1180$ &$+25\%$
\\
 $\sqrt{\langle r^2_{E,S}\rangle}$ (fm)  & $0.79 $ &  $0.87$      &$+21\%$
\\
 $g_A$ & $1.25 $ & $0.98$&$-21\%$
\\
 $\mu_p-\mu_n$ & $4.7 $ & $3.7$&$-22\%$
\\
 $\Gamma(\omega\rightarrow \pi\gamma)$ & $0.75$   & $0.82$
           &$+10\%$
            \\
 $\Gamma(\omega\rightarrow 3\pi)$ & $7.5$   & $7.1$      &$-6\%$
            \\
 $\Gamma(\rho\rightarrow \pi\gamma)$ & $0.068$   & $0.072$          &$+5\%$
 \\
 $\Gamma(\omega\rightarrow \pi\mu\mu)$ & $8.2\cdot 10^{-4}$   & $7.4\cdot 10^{-4}$      &$-9\%$
 \\
 $\Gamma(\omega\rightarrow \pi e e )$ & $6.5\cdot 10^{-3}$   & $7.4\cdot 10^{-3}$         &$+14\%$
           \\
\hline
\end{tabular}
\caption{Global fit of mesonic and baryonic  physical quantities. 
Masses, decay constants and widths  are given in MeV.
The RMS error  of the fit is $16\%$. 
Physical masses have been used in the kinematic factors of the partial decay  widths.}
\label{pretotal}
\end{center}
\end{table}

The global fit gives $1/L\simeq 350$ MeV, $M_5L\simeq 0.017$ and 
 $\alpha\simeq 0.9$ ($\gamma\simeq 1.27$).
As expected, the value of $\alpha$ from the fit  is   close to 1, as predicted
in the  large-$N_c$ limit. 
Also it is worth noticing that  the value of $M_5L$ comes to be  quite close 
to the prediction arising from matching the current-current vector correlator 
to QCD at large momentum  \cite{Erlich:2005qh,Rold:2005}, $M_5L=N_c/(24\pi^2)\simeq  0.013$.
The RMS error of the fit is $16\%$.
From the fit one can see that mesonic quantities are better predicted than the
baryonic ones.
Indeed, mesonic quantities alone give a fit with a 
RMS error around $10\%$, of order of the experimental errors.

\section{Conclusions}

Using a simple holographic model for QCD, we have been able to compute
the  properties of the baryons.
We have seen that the presence of the CS term, needed to reproduce the QCD anomaly,
is crucial to stabilize the size of the baryon at around the GeV.
The contribution of this CS term to the mass of the baryon 
is as important as the leading $F^2$ term, implying
the need for a fully numerical analysis of the non-linear solitonic configuration,
not carried out before in the literature.

We have calculated the axial coupling, the vector magnetic moment and the isoscalar
electric charge radius of the nucleus 
as a function of the 3 free parameters of the model.
These predictions have shown a good agreement with data -see Table~\ref{preba}.
Our approach can be extended to calculate other baryonic physical quantities
that we leave for future work.

We have also seen that the CS term is  responsible for  anomalous processes involving
an odd number of pions, some of which we have explicitly calculated (see Table~\ref{preano})
and showed to have an excellent agreement with the experimental data.
Finally, we have done a combined fit of the mesonic and baryonic
quantities predicted by this holographic model (Table~\ref{pretotal})
that have shown  an agreement with experiments of  $16\%$.

Some final comments on the important issue of calculability, {\it i.e.} on the dependence 
of our results on higher-dimensional operators. These operators are suppressed by the 
cut-off of the theory $\Lambda_5$. 
Naive dimensional  arguments say that 
the maximal value of $\Lambda_5$ is determined by the scale
at which   loops   are of order  of tree-level effects.
Computing loop corrections to the $F^2$ operator which arise from 
the $F^2$ term itself, one gets  $\Lambda_5\sim 24\pi^3 M_5$. 
Nevertheless, one gets a lower value for $\Lambda_5$ from the CS term
due to the $N_c$ dependence of $\gamma$.
Indeed, at the one-loop level, the CS term gives a contribution of order $M_5$
to the $F^2$ operator for
$\Lambda_5\sim 24\pi^3M_5/N_c^{2/3}$.
Even though  the cut-off scale  lowers due to the presence of the CS term,  we can still have, in the large-$N_c$ limit,
a 5D weakly coupled theory where higher-dimensional operators are suppressed. 
 For  $M_5\sim N_c/(16\pi^2L)$, as expected from the AdS/CFT correspondence, one has
in fact $\Lambda_5L\sim  N_c^{1/3}\rightarrow\infty$. 
The same  factor protects
the physics of the 5D skyrmions 
which have size $L$.

\section*{Acknowledgments}

A.P. thanks Rafel Escribano for discussions.
The work  of A.P. was  partly supported   by the
FEDER  Research Project FPA2005-02211
and DURSI Research Project SGR2005-00916.



\begin{thebibliography}{99}
  

\bibitem{'tHooft:1973jz}
  G.~'t Hooft,
  Nucl.\ Phys.\  B {\bf 72} (1974) 461.


\bibitem{Skyrme:1961vq}
  T.~H.~R.~Skyrme,
  Proc.\ Roy.\ Soc.\ Lond.\  A {\bf 260} (1961) 127.

\bibitem{Adkins:1983ya}
  G.~S.~Adkins, C.~R.~Nappi and E.~Witten,
  Nucl.\ Phys.\  B {\bf 228} (1983) 552.

\bibitem{Meissner:1987ge}
  U.~G.~Meissner,
  Phys.\ Rept.\  {\bf 161} (1988) 213.


\bibitem{Igarashi:1985et}
    Y.~Igarashi, M.~Johmura, A.~Kobayashi, H.~Otsu, T.~Sato and S.~Sawada,
  Nucl.\ Phys.\  B {\bf 259} (1985) 721.


\bibitem{Hata:2007mb}
  H.~Hata, T.~Sakai, S.~Sugimoto and S.~Yamato,
  arXiv:hep-th/0701280.


\bibitem{Pomarol:2007kr}
  A.~Pomarol and A.~Wulzer,
  JHEP {\bf 0803} (2008) 051.

\bibitem{Erlich:2005qh}
  J.~Erlich, E.~Katz, D.~T.~Son and M.~A.~Stephanov,
  Phys.\ Rev.\ Lett.\  {\bf 95} (2005) 261602.

\bibitem{Rold:2005}
  L.~Da Rold and A.~Pomarol,
  Nucl.\ Phys.\  B {\bf 721} (2005) 79; JHEP {\bf 0601} (2006) 157.

\bibitem{Hirn:2005nr}
  J.~Hirn and V.~Sanz,
  JHEP {\bf 0512} (2005) 030.


\bibitem{Nawa:2006gv}
  K.~Nawa, H.~Suganuma and T.~Kojo,
  Phys.\ Rev.\  D {\bf 75} (2007) 086003;
  Prog.\ Theor.\ Phys.\ Suppl.\  {\bf 168} (2007) 231.


\bibitem{Hong:2006ta}
  D.~K.~Hong, T.~Inami and H.~U.~Yee,
  Phys.\ Lett.\  B {\bf 646} (2007) 165;
  D.~K.~Hong, M.~Rho, H.~U.~Yee and P.~Yi,
  Phys.\ Rev.\  D {\bf 76} (2007) 061901;
  Phys.\ Rev.\  D {\bf 77} (2008) 014030.

\bibitem{Hata:2008xc}
  H.~Hata, M.~Murata and S.~Yamato,
  arXiv:0803.0180 [hep-th];
  K.~Hashimoto, T.~Sakai and S.~Sugimoto,
  arXiv:0806.3122 [hep-th].

\bibitem{Sakai:2004cn}
  T.~Sakai and S.~Sugimoto,
  Prog.\ Theor.\ Phys.\  {\bf 113} (2005) 843.

\bibitem{Gubser:1998bc}
S.~S.~Gubser, I.~R.~Klebanov and A.~M.~Polyakov,
Phys.\ Lett.\ B {\bf 428} (1998) 105.

\bibitem{Witten:1998qj}
  E.~Witten,
  Adv.\ Theor.\ Math.\ Phys.\  {\bf 2} (1998) 253.

\bibitem{Hill:2006wu}
  C.~T.~Hill,
  Phys.\ Rev.\  D {\bf 73} (2006) 126009.

\bibitem{Panico:2007qd}
  G.~Panico and A.~Wulzer,
  JHEP {\bf 0705} (2007) 060.



\bibitem{Weinberg:2006rq}
See,  for example,  E.~J.~Weinberg and P.~Yi,
  Phys.\ Rept.\  {\bf 438} (2007) 65.


\bibitem{Witten:1979kh}
  E.~Witten,
  Nucl.\ Phys.\  B {\bf 160} (1979) 57.





\bibitem{comsol}
See http://www.comsol.com.






  
  
\bibitem{Yao:2006px}
  W.~M.~Yao {\it et al.}  [Particle Data Group],
  J.\ Phys.\ G {\bf 33} (2006) 1.
  
  
  \bibitem{Sakai:2005yt}
  T.~Sakai and S.~Sugimoto,
  Prog.\ Theor.\ Phys.\  {\bf 114} (2006) 1083.
  
  
\bibitem{Kaymakcalan:1983qq}
  O.~Kaymakcalan, S.~Rajeev and J.~Schechter,
  Phys.\ Rev.\  D {\bf 30} (1984) 594.
  
\bibitem{Grigoryan:2008up}
  H.~R.~Grigoryan and A.~V.~Radyushkin,
  arXiv:0803.1143 [hep-ph].
      
  
\end{thebibliography}
\end{document}